\documentstyle[11pt,aasms4,natbib]{article}

\def\eg{{\it e.g., \,}}

\vspace{14cm}
\includegraphics{Bgrad.ps}

\begin{document}
\title{Grey Milky Way Extinction from SDSS Stellar Photometry}

\author{Evgeny Gorbikov \& Noah Brosch}
\affil{The Wise Observatory and
the Raymond and Beverly Sackler School of Physics and Astronomy,
the Faculty of Exact Sciences,
Tel Aviv University, Tel Aviv 69978, Israel}


\begin{abstract}
We report results concerning the distribution and properties of galactic extinction at high galactic latitudes derived from stellar statistics using the Sloan Digital Sky Survey (SDSS). We use the classical Wolf diagram method to identify regions with extinction, and derive the extinction and the extinction law of the dust using all five SDSS spectral bands. We estimate the distance to the extinguishing medium using simple assumptions about the stellar populations in the line of sight.

We report the identification of three extinguishing clouds, each a few tens of pc wide, producing 0.2--0.4 mag of $g'$-band extinction, located 1--2 kpc away or 0.5--1 kpc above the galactic plane. All clouds exhibit grey extinction, i.e., almost wavelength-independent in the limited spectral range of the SDSS.
We discuss the implication of this finding on general astrophysical questions.
\end{abstract}


\section{Introduction}

The Milky Way (MW) extinction was studied since the beginning of the 20$^{th}$ century to determine the wavelength dependence of the galactic extinction (the galactic extinction law), and to map the extinction in the different directions of the Galaxy. \cite{SCH75} showed that for a wide spectral range the MW extinction law is $A_{\lambda}\sim\lambda^{-1}$, thus the extinction
curve can be parameterized by a single variable $R_V$, the total-to-selective extinction ratio, defined as
\begin{equation}
R_V = \frac{A_V}{E(B-V)}~,
\label{eq:TTSEratio}
\end{equation}
where $E(B-V) = A_B - A_V$ is the colour excess or the reddening. $R_V$ varies in different directions of the MW \citep{MAT90}, especially toward dense dust clouds. The mean value $<R_V>\cong3.14\pm0.10$ \citep{SCH75} and the typical galactic extinction law, that behaves as $A_{\lambda}\sim\lambda^{-1}$, are only MW averages.
Wavelength-independent extinction $A_\lambda \sim$ Const, which is produced by large dust particles with sizes $a \ge \lambda$, is called
'grey' extinction usually. For grey extinction the reddening converges to zero, and therefore $R_V \gg 3.1$.
Knowledge of the interstellar extinction measurements and mapping is important for stellar studies and also for extra-galactic astronomy. Extinction maps are needed to correct the galactic and extra-galactic observations for the MW extinction.

The first photometric method for measuring the extinction was proposed by \cite{WOL23} and uses ``Wolf diagrams''. The method is based on the assumption that dust located between the object and the observer reduces the brightness of the object, thus the apparent magnitude $m_{\lambda}$ of the object increases by $A_{\lambda}$ (the extinction in magnitude units). In presence of extinction, the distance modulus is:
\begin{equation}
m_\lambda - M_\lambda = -5 + 5\log\left(\frac{d}{1 pc} \right) + A_{\lambda}~,
\label{eq:dist_modulus3}
\end{equation}
where $d$ is the distance between the object and the observer.

\begin{figure*}[ht!]
\vspace{6cm}
\includegraphics{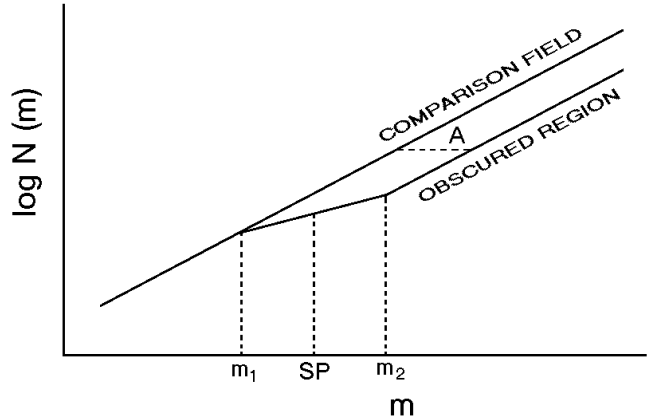}
\caption{%
Schematic Wolf diagram: star count distribution of an extinguished region and of a reference field covering the same area, as a function of magnitude. The distributions are assumed to be the same, but due to the presence of the obscuring material, the distribution of the extinguished region is shifted by $A$ magnitudes. The magnitudes m1, m2 and the splitting point SP are defined in Section 2.2.\label{Fig:wolf_example}
}
\end{figure*}

It is possible to derive $A_\lambda$ by comparing the cumulative star count distribution in an extinguished region with that of a dust-free reference field. The two distributions are assumed to be identical, but the dust extinction shifts the distribution of the extinguished region, as shown in Figure \ref{Fig:wolf_example}. Since the stellar distribution in the MW is not uniform, the distribution slopes of the two regions may change with the apparent magnitude, but will remain parallel. The star count distribution changes with the galactic location, thus the two regions must be sufficiently close to reduce these differences.

The ``Wolf diagrams'' method was widely applied for extinction measurements between the 1930s and 1960s (e.g., Bok 1956). As recently as 1996 the method used star counts performed by eye (Andreazza \& Vilas-Boas 1996). The availability of digital catalogs with positions, magnitudes, and sometimes colors, eased significantly this tedious and error-prone task (\textit{e.g.}, \citealt{CHA99A,CHA99B,NIK01,KRA03,FRO05}). \cite{CHA99A,CHA99B} used USNO catalog star counts and the Wolf diagrams method to derive an all-sky extinction map and detailed extinction maps of giant molecular clouds. \cite{FRO05} used 2MASS cumulative star counts to obtain detailed extinction maps of the Galactic plane. 

The implementation of the method has two main problems. The first is the proper choice of the unextinguished reference field. The second is the stellar density and the luminocity function variations, which originate from the galactic structure itself. The latter problem was solved by \cite{CHA99A} using an adaptive area algorithm, which changed the analyzed region area to keep the star counts constant.

The use of this method is still limited due to difficulties in the automatization of the algorithm, the large amounts of CPU and manpower requirements, and the lack of a deep optical or UV photometric stellar surveys covering both the regions of high and low galactic extinction. One advantage of the Wolf method is that it allows the estimation of the distance to the extinguishing cloud as the typical distance to the stars at the magnitude corresponding to the break in the cumulative star count distribution of the extinguished region. The other advantage of this method is that it deals directly with the extinction and not with the reddening, and is therefore sensitive not only to ``normal'' dust, but also to dust that produces no reddening.

Another photometric method developed in early 1990's deals with star colours instead of magnitudes. Since for ``normal'' dust the extinction $\propto\lambda^{-1}$, extinction affects different colours differently. The extinction can be extracted from the colour-colour diagram of the extinguished region, since the extinguished stars are shifted in the colour-colour diagram by the colour excesses. Implementations of this method were by \cite{LAD94}, \cite{LOM01} and \cite{SAL08}.

The advantages of the colour-colour method are simplicity, robustness and full automation, while the disadvantages are a strong dependence of the results on statistical and systematic errors in the photometric data, and the assumption of a wavelength-dependent extinction.

Dust and gas are well mixed in the MW; one can therefore measure the gas column density associated with a dust cloud and estimate the extinction using a gas-to-dust ratio. 
The main disadvantage of the gas-based method is that the quantities are connected to dust properties only through average ratios, and thus may differ for particular locations. \cite{GOO08} compared three different methods of extinction estimation for the Perseus star-forming region: mapping using near-IR colour excesses, dust far-IR emission, and mapping the integrated intensity of the $^{13}$CO emission. While all the methods produced morphologically similar maps, the gas-based extinction values deviated greatly from the two others.

The obvious advantage of the gas-based method is the data reduction simplicity.
The convenience of this method led to the production of the first all-sky galactic extinction map. \cite{BUR78} solved the major problems of the reddening biases and varying gas-to-dust ratio by applying a complicated algorithm to the HI spectral data. They combined 21-cm measurements with galaxy counts in the direction of 49 globular clusters, 84 RR Lyrae stars, and two early-type stars, and produced an all-sky galactic extinction map that had 10--20\% accuracy and 13 square degrees spatial resolution for regions with $|b|>10^{\circ}$ and $z$ distances larger than 300 pc (\citealt{BUR82,BUR84}). From the early 1980's to the end of the 1990's the Burstein \& Heiles maps were the only tool for the galactic extinction correction of galactic and extra-galactic observations.

%

The last method uses the far infra-red (FIR) dust emission that is dominated by thermal emission from dust grains.
%
\citet[][hereafter SFD]{SCH98} combined IRAS and COBE/DIRBE measurements to create a relatively high resolution (based on IRAS) and well-calibrated (based on COBE) all-sky galactic extinction map. The authors used 100 and 240 micron COBE/DIRBE measurements to create a map of the galactic dust temperature. The combination of this map with IRAS high-resolution 100 micron intensity map provided an all-sky galactic reddening map with 16\% accuracy and 1/8 $\times$ 1/8 square degrees spatial resolution.

The SFD maps still remain the most powerful and commonly used tool for the galactic extinction correction (\eg the foreground galactic extinction estimates in the NASA/IPAC Extragalactic Database), but cannot be used to determine the distance to a dust cloud since they provide only a cumulative effect of FIR dust emission. It is clear that the SFD maps tend to overestimate the extinction. \cite{ARC99} pointed out that the SFD maps overestimate the reddening by a factor of 1.3-1.5 in regions of smooth extinction with $A_{V} > $ 0.5 mag. \cite{YAS07} found that this deviation is connected with a degeneracy between the 100 and 240 micron measurements, which does not allow a proper black-body spectrum fitting to the data. \cite{YAH07} found that the FIR emission from galaxies affects the SFD reddening maps that tend to overestimate the extinction in regions with background galaxies. Here we shall demonstrate that SFD maps can in some cases also underestimate the extinction.

%
Most methods listed above are not sensitive to grey extinction, since they deal with reddening and not with the extinction, or since they use average MW ratios. The exceptions are the Wolf diagrams and the FIR emission maps. The SFD maps can determine the presence of grey dust, but do not distinguish between ``normal'' and grey dust. This leaves the Wolf diagrams as the only method useful for detecting both ``normal'' and grey dust.

Large 1 $\mu$m dust grains, which produce grey extinction in the visible part of the EM spectrum, are only a tiny fraction of the MW grain size distribution (\citealt{MAT77,KIM94}). Large quantities of grey dust grains could be found only in places where physical processes selectively support the production or survival of large-size particles.

Grey extinction was detected by Sitko et al. (1984), Dunkin \& Crawford (1998), Barge \& Viton (2003), Patriarchi et al. (2003) and Skorzynski et al. (2003) in circumstellar disks of some MW stars. Patriarchi et al. (2003) pointed out the connection of grey extinction with some local features in the Carina association. The detections of Patriarchi et al. (2003) and Skorzynski et al. (2003) were questioned by \cite{MAI05}, who argued that grey extinction was determined mistakenly due to an incorrect treatment of the Lutz-Kelker bias. The presence of grey dust in a circumstellar disk was modeled and explained by \cite{MEN99} as a result of an initial phase of planetary system formation, when large particles coagulate from small grains. 
Grey extinction was also discovered in high-redshift GRB host galaxies by \cite{SAV03}, \cite{SAV04}, \cite{STR04}, \cite{STR05}, \cite{CHE06}, \cite{LII08} and \cite{PER08}. \cite{SAV03} proposed that small dust grains are preferentially destroyed by GRBs and only the large grains survive. 

Greyish extinction (weakly wavelength-dependent) was found also in supernova remnants by Todini \& Ferrara (2001), Bianchi \& Schneider(2007), Nozawa et al. (2008). The origin of the greyish extinction in supernova remnants is probably the same as in the GRB host galaxies: small dust grains being selectively destroyed by the supernova outburst.

%
The main purpose of our study was to detect dust clouds at high galactic latitude and evaluate their properties. The SFD map predicts that in the studied region the extinction is $A_{g'} < 0^m.4$. Based on this, one would not expect to find high-extinction dust clouds in this region. We used the Wolf diagram method, measuring the extinction directly from the stellar statistics. Comparing our results to the SFD maps, we can confirm or reject the SFD predictions in this area. We identified three rather compact regions that show ``grey'' or greyish extinction.

We also intended to detect possible significant galactic extinction in the direction of galaxies of the Hubble Space Telescope Key Project (hereafter KP) that aimed to determine H$_0$ to $\pm 10\%$ \citep{FRE01}. If the extinction in the direction of the KP galaxies would not have been properly estimated, this could have induced a systematic error in the H$_0$ measurement. The KP galaxies were chosen so that the SFD extinction prediction in their directions would be small ($A_g < 0^m.1$). Our results showed no significant extinction for three KP galaxies (NGC 4321, NGC 4548, NGC 4535) located in the region of study (Figure \ref{Fig:SFD_total_equ}).
    
\section{Data and reductions}

\begin{figure*}[ht!]
\vspace{11cm}
\includegraphics{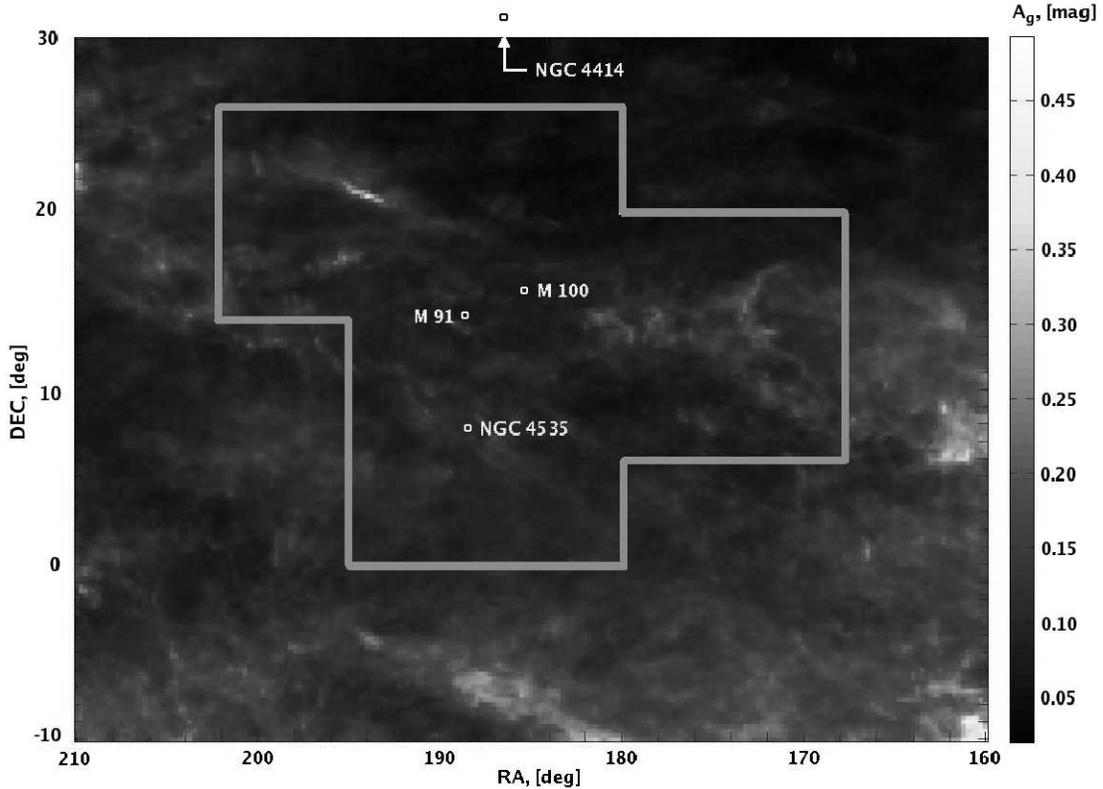}
\caption{%
The $g'$ band SFD extinction map for the Virgo cluster region plotted in equatorial coordinates. The grey solid line
shows the border of our database. White circles indicate the locations of four galaxies from the
HST Key Project.
   \label{Fig:SFD_total_equ}}
\end{figure*}

We used data from the SLOAN Digital Sky Survey (SDSS) DR6 with technical details given in \cite{ADE08}.
They include a list of objects recognized as stars by the SDSS pipeline with their equatorial coordinates, the apparent magnitudes in five SDSS bands -- $u'g'r'i'z'$ and the photometric errors in each band.
The SDSS limiting magnitude ($u'_{lim}=22.0,$ $g'_{lim}=22.2,$ $r'_{lim}=22.2,$ $i'_{lim}=21.3$ and $z'_{lim}=20.5$) is defined as the magnitude of the faintest point source with a probability of detection repeatability of at least 95\%. Considering this, we excluded from our study all the stars that do not satisfy the SDSS magnitude limits in at least one of the bands. 
The region studied here is plotted in Figure~\ref{Fig:SFD_total_equ} and covers $\sim650$ square degrees of solid angle in the Virgo and Coma Berenices constellations. The approximate borders of the area are $\alpha = 11.^{h}2$ to $\alpha = 13.^{h}5$ and $\delta = 0^\circ$ to
$\delta = 26^\circ$.

\subsection{Tiling and the ``Bull's eye'' algorithm}

We mentioned in the Introduction specific details of the Wolf diagrams method and some shortcomings of this method. We use the method here and solve the problems of the reference field and of galactic structure variations by using galactic coordinates and applying an algorithm called ``Bull's eye''.

\begin{figure*}[ht!]
\begin{center}
\vspace{11cm}
\includegraphics{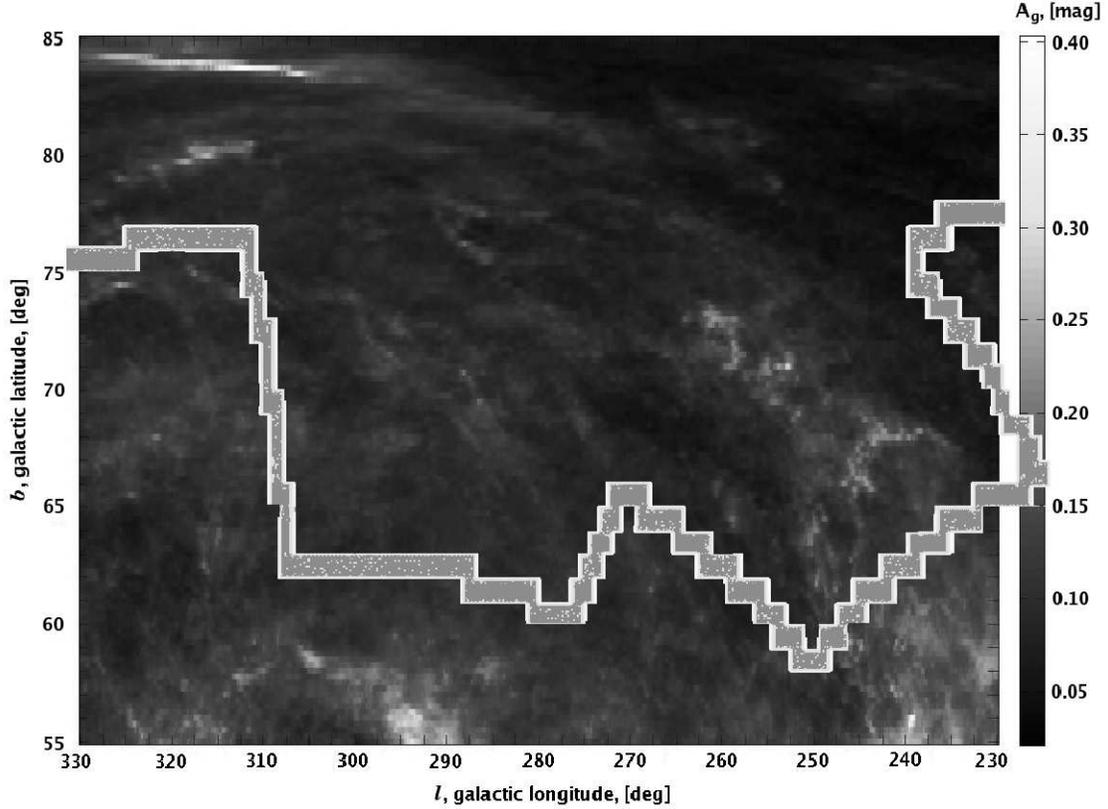}
\end{center}
\caption{%
The $g'$ band SFD extinction map of the Virgo cluster region displayed in galactic coordinates. The grey solid
line shows the rejected cells at the border of the database. The sample region is located above the grey line.
   \label{Fig:SFD_total_gal}}
\end{figure*}

The coordinates of all the stars in the original database were transformed to the galactic coordinate system, and the entire region was divided into one square degree cells. We rejected 433 cells at the border of the region, since parts may actually be beyond the region defined in celestial coordinates, and retained 2632 cells, each with $\sim$1000 to $\sim$5000 stars. Figure~\ref{Fig:SFD_total_gal} shows the sky coverage in galactic coordinates.

%

The selection of a reference field for each cell, and the averaging of star count differences caused by the galactic structure, were obtained using the ``Bull's eye'' algorithm.
We defined the cumulative star count distribution in cells within a
small area centered on the extinguished region, and compared it with that of cells within a
region surrounding the selected area. The reason for choosing two concentric or quasi-concentric regions for the analyzed and for the reference field is to eliminate differences in stellar density and luminosity distribution in different cells due to a change of viewing direction within the Milky Way. Changes in the projected stellar density and the luminosity distribution with (\textit{l}, \textit{b}) are expected because the MW stars are not distributed uniformly. By selecting a surrounding concentric comparison area around the extinguished region we average over such effects.

We assumed that the cells within the surrounding region are not extinguished. 
This implies that we assumed a small and relatively compact extinguishing cloud, well-separated from its surroundings. In the worst case, if the cells within the comparison region would also be extinguished, we would find only the
differential extinction between the target and the comparison region. The distributions were normalized with respect to the areas of the two regions, so that only projected stellar densities per square degree were compared.




Our stellar selection program allows
choosing particular cells from the database and defines the surrounding comparison cells. It is therefore possible to choose regions with arbitrary shapes, not necessarily having circular symmetry. All the reference fields assumed to be unextinguished in this study were chosen to be within two degrees from the analyzed regions.

\subsection{Determining extinction and distance}
A difference between the cumulative star count distribution of the reference field and that of the extinguished region causes a horizontal shift of the latter by the extinction $A$ (see Figure 1). In practice, the distributions begin to diverge from a location at $m_1$. The slopes of the two distributions return to the same value at $m_2$, and from that point on the cumulative distributions remain parallel. The average magnitude between the two points $m_1$ and $m_2$ is called here the ``splitting point'' (SP). The estimation of the extinction $A$ can be performed by averaging the distance on the magnitude axis between the two graphs, as done by \cite{WOL23}. 
We defined the extinction for each magnitude bin starting at $m_2$ and averaged the obtained extinction values. The error in extinction was estimated as the standard deviation of the extinction values for each magnitude bin from the average value. 
To compare the cumulative star count distributions in two regions we performed a $\chi^2$ Goodness-of-Fit test.


We determined the distance to the extinguishing medium as the typical distance to the stars located at the SP. This was determined from
the apparent magnitude $m_\lambda$ and the absolute magnitude \textit{$M_\lambda$} of each star.
To find the distance for each star we used only the SDSS $g'r'i'$ data. We transformed the $(g' - r')$ and $(r' - i')$ colours into $(B - V)$ and $(R_c - I_c)$ respectively, and determined the apparent magnitude $m_V$ using the \cite{FUK96} relations. We used the HR diagram data for main sequence stars from \cite{COX00} and a cubic spline interpolation to derive \textit{$M_V$} from the stellar colours. The $(g' - r')$ and $(r' - i')$ colours yielded two distance estimates for each star, marked as $D_{(g - r)}$ and $D_{(r - i)}$. We stress that these distances are valid only for main sequence stars, which represent $\gtrsim90\%$ of stars (\citealt{MEN79,JUR08}).

For further distance measurements we used only one distance estimate for each star, provided that the relative difference between the two estimates is small:
\begin{equation}
\left\vert \varepsilon(D) \right\vert = \left\vert \frac{D_{(g - r)} - D_{(r - i)}}{D_{(g - r)} + D_{(r - i)}} \right\vert \leq 0.1
\label{eq:criterion}
\end{equation}
Stars with larger $\varepsilon(D)$ estimates have great distance differences due to unusual colour relations, thus they probably are not main-sequence stars or may be binaries. The number of stars with large $\varepsilon(D)$ estimates is consistent with the number of non-main sequence stars \citep{MEN79,JUR08}. The stars that do not fit the SDSS limiting conditions ($\sim20\%$ of the total sample), the stars with large relative difference between the distance estimates($\sim9-10\%$) and the stars too red for the \cite{FUK96} relations ($\sim7\%$) were rejected from further processing. The rejeted stars make up to $\sim40\%$ of the SDSS stars in each sky tile.

We estimated the extinction and the distance to the extinguishing medium in an alternative way with a simple model of a thin uniform dust screen with extinction $A_\lambda$, located at a distance $d$. We modeled the star count distribution in the direction of the extinguished region by applying this thin screen, with parameters $(d, A_\lambda)$, to the star count distribution of the reference field. We assumed that every star more distant than $d$ is extinguished, while all stars nearer than $d$ retain their apparent magnitudes unchanged. We used the stellar distances derived previously. We estimated the best-fit distance and extinction for each dust cloud $(d_{min}, A_{min})$ by minimization of the fit in the $(d, A)$ parameter space. We used the downhill simplex algorithm to minimize the $\chi^2$ Goodness-of-Fit value between the distribution of the extinguished region and the modeled distribution.
%
The errors of the fit are determined with the ``constant $\chi^2$ boundaries'' method described in Chapter 15.6 of \cite{Press1992}.

\section{Analysis and results}

\begin{figure*}[ht!]
\begin{center}
\vspace{9cm}
\includegraphics{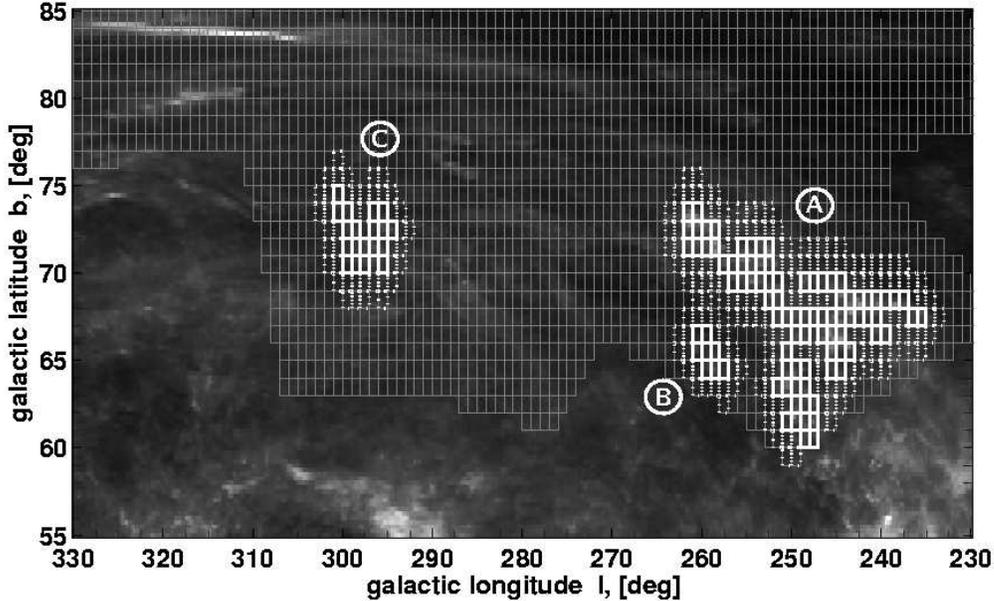}
\end{center}
\caption{%
Regions found to show no extinction where the SFD map predicted significant extinction. Grey rectangles represent the one-square-degree cells of the database. White solid rectangles denote the cells of the tested regions, white dashed rectangles denote cells of the reference fields.
\label{Fig:no_ext}}
\end{figure*}

\begin{figure*}[ht!]
\begin{center}
\vspace{12cm}
\includegraphics{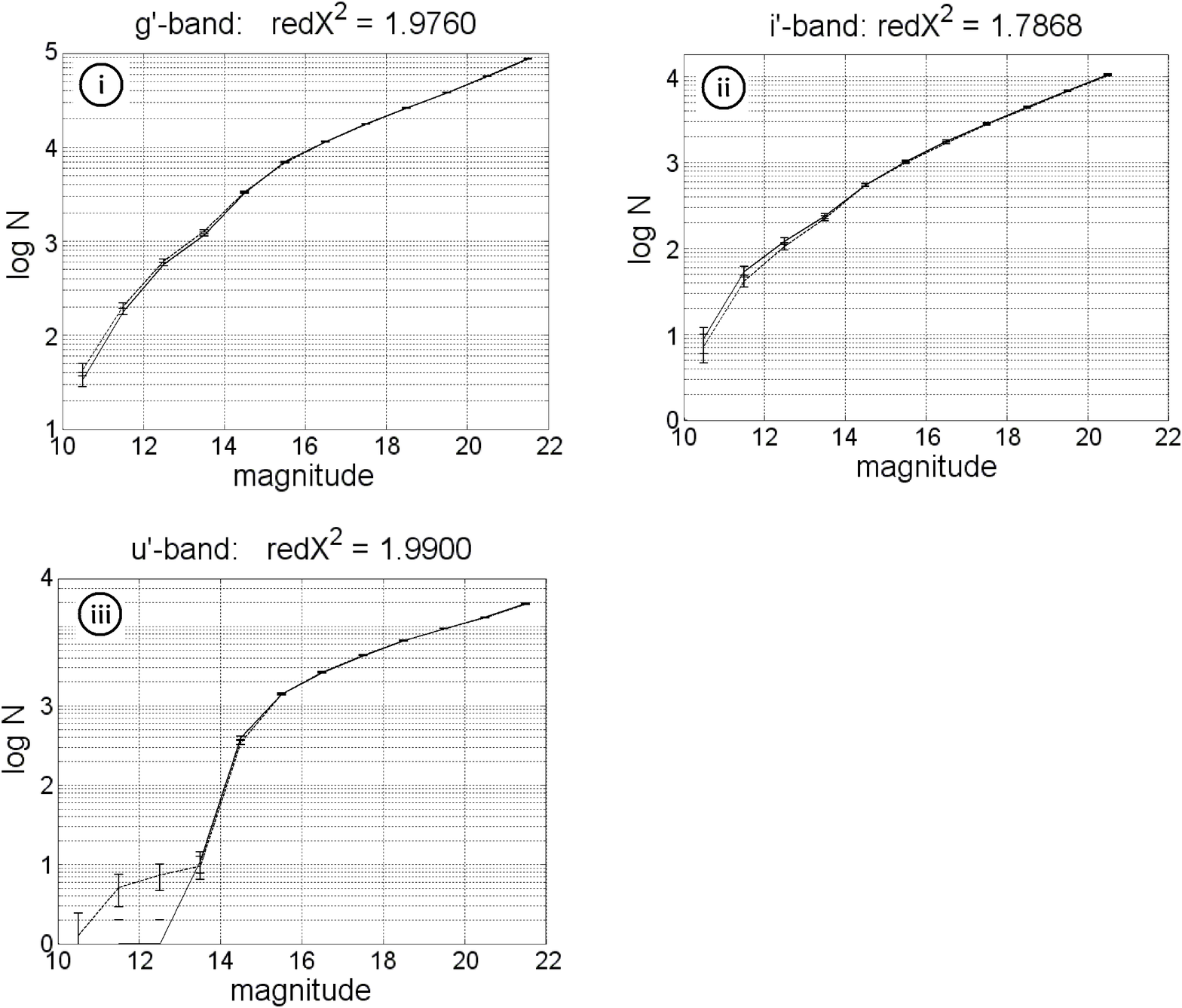}
\end{center}
\caption{%
Examples of Wolf diagrams for the regions A, B and C. Solid lines are for the star count cumulative distribution in the analyzed regions, dashed lines are for the star count cumulative distribution in the reference fields. The title of each panel shows the value of the reduced $\chi^2$ GoFT between the distributions. Panel (i) shows the $g'$ band Wolf diagram for the region A, panel (ii) shows the $i'$ band Wolf diagram for the region B, panel (iii) shows the $u'$ band Wolf diagram for the region C. The behavior of all the SDSS bands is similar. 
\label{Fig:ABCWolf}}
\end{figure*}

We selected three regions A, B  and C from the surveyed area where the SFD predictions were large and checked their Wolf diagrams. These regions are shown in Figure \ref{Fig:no_ext} and are centered on $(l, b)$: $(248^\circ, 67^\circ)_A$, $(260^\circ, 65^\circ)_B$ and $(297^\circ, 73^\circ)_C$ respectively. The areas of the regions are $\Omega_A\sim35.9$, $\Omega_B\sim3.2$ and $\Omega_C\sim5.7$ square degrees respectively. The SFD predictions for regions A and B are $A_{g'}\sim0.^m20$, and $A_{g'}\sim0.^m25$ for region C. Examples of selected Wolf diagrams for the regions A, B and C are shown in Figure \ref{Fig:ABCWolf}. The errors of star count distributions are defined as:
\begin{equation}
\Delta N(m) = \sqrt{N}/\sqrt{j}~,
\label{eq:deltaN}
\end{equation}
where $\sqrt{N}$ is the Poisson error of star counts and $j$ is the number of cells, which represents the area of the analyzed region or the reference field.

The three regions do not show significant extinction, despite the SFD prediction. We can set an upper limit to the extinction in these regions as $A_{g'} < 0.^m1$. A number of explanations can be proposed to understand the discrepancy:

\begin{itemize}
\item There may be extinction in these areas, but it is much smaller than the SFD prediction, thus the SFD map overestimates the extinction. This was mentioned by \cite{ARC99}, \cite{YAS07} and \cite{YAH07}.
\item The extinction source in these areas may be very distant, perhaps even extragalactic, and the SDSS does not reach deep enough to allow its detection using a stellar statistics analysis. We estimate the limiting distance of our study to be of order 10 kpc, as the typical distance of the stars at the SDSS limiting magnitude in the region of study.
\item The determination of the extinction in this study is performed through a comparison of the selected region and its surroundings. If the reference field itself is extinguished, no extinction can be detected using the Wolf diagrams method. We can probably rule out this possibility, since the SFD prediction for the reference fields of all three regions is $A_{g'} \sim 0.^{m}1$ indicating that, from a thermal IR point of view, there may only be a small amount of dust in these regions. 
\end{itemize}

\begin{figure*}[ht!]
\begin{center}
\vspace{11cm}
\includegraphics{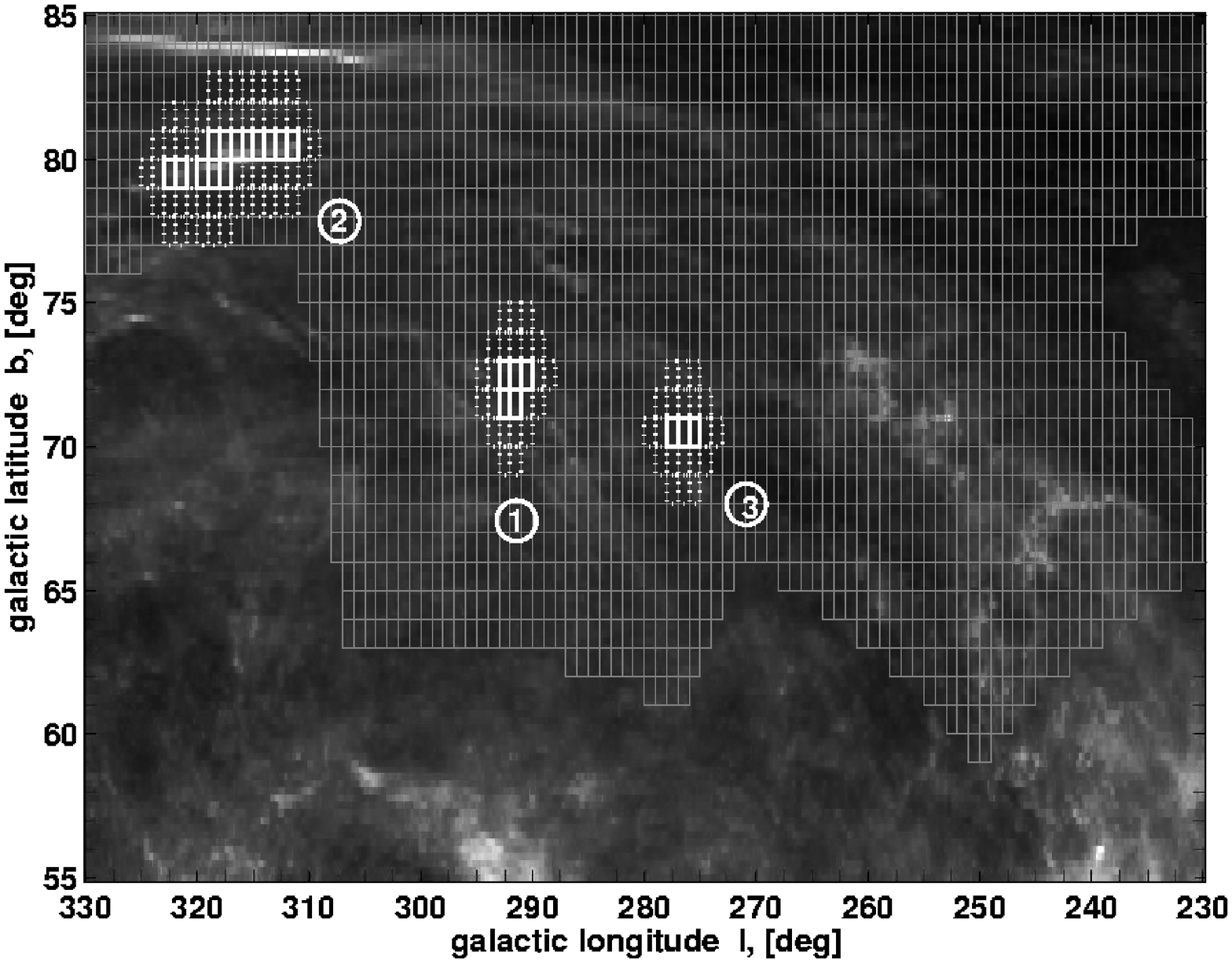}
\end{center}
\caption{%
Extinguished regions found in this study. White solid rectangles denote cells of the extinguished regions, white dashed rectangles mark cells of the reference fields.
\label{Fig:ext_all}}
\end{figure*}

\begin{figure*}[ht!]
\begin{center}
\vspace{15.5cm}
\includegraphics{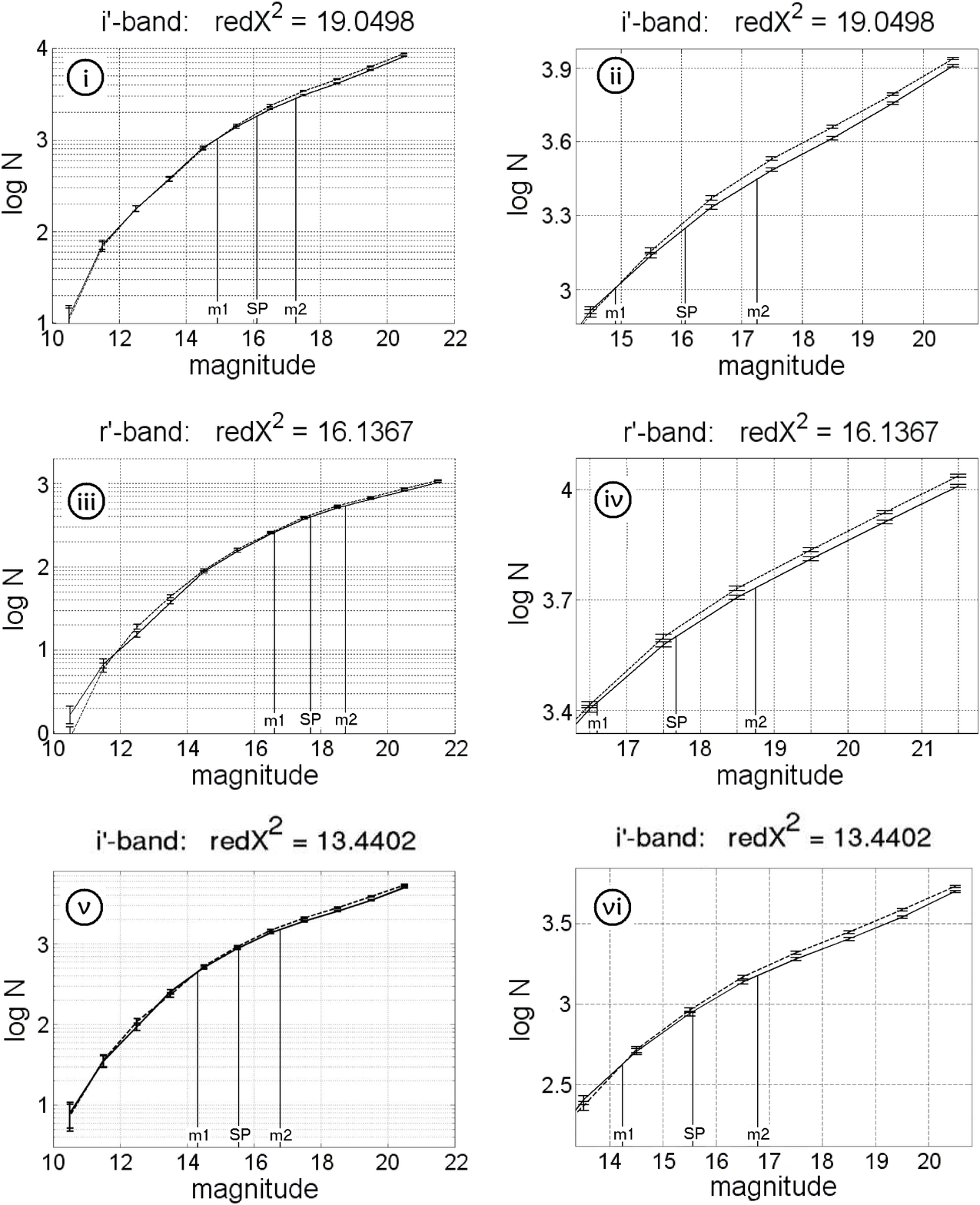}
\end{center}
\caption{%
Examples of Wolf diagrams for the regions 1, 2 and 3. The notation is as in Figure \ref{Fig:ABCWolf}. Panel (i) shows the $i'$ band Wolf diagram for the region 1, panel (iii) shows the $r'$ band Wolf diagram for the region 2, panel (v) shows the $i'$ band Wolf diagram for the region 3. Panels (ii), (iv) and (vi) are the enlarged plots of panels (i), (iii) and (v) respectively. m1, m2 and SP denote respectively the magnitudes m1, m2 and the splitting point defined in Section 2.2.
\label{Fig:123Wolf}}
\end{figure*}

We also checked the locations of three HST Key Project galaxies that are included in the survey region and were chosen to be in locations with small extinction: $A_{g'} < 0.^{m}1$. We did not find significant amounts of extinction at their locations.

Finally, we scanned the entire surveyed region in $g'$ band using the ``Bull's eye'' method described above. We performed the scan cell-by-cell, taking the central cell as analyzed region and the surrounding cells within two degrees from the central cell as a reference field. We used the $\chi^2$ GoFT on the normalized star count cumulative distributions in order to detect cells with significant deficiency of faint stars comparatively to the reference field. We identified three locations showing a deficiency of faint stars. We examined these regions using the other SDSS bands data and found that they showed a significant deficiency of faint stars in all SDSS bands. We combined the data for the extinguished cells in each of the regions to obtain more accurate Wolf diagrams. The three extinguished regions are shown in Figure \ref{Fig:ext_all} and their Wolf diagrams are shown in Figure \ref{Fig:123Wolf}. The errors of star count distributions are defined by Equation (\ref{eq:deltaN}).

For each region we calculated the extinction in each SDSS band and compared it with that obtained from the SFD prediction using the extinction ratios for the SDSS bands derived by \cite{FUK04} and \cite{YAS07}, namely $A_{u'} : A_{g'} : A_{r'} : A_{i'} : A_{z'} = 5.2 : 3.8 : 2.8 : 2.1 : 1.5$. We used the accuracy of the SFD map, 16\% of measured reddening value \citep{SCH98}, to calculate an error of the predicted extinction. Below we present the results for each of the regions together with a short discussion.

\subsection{First extinguished region}

\begin{figure*}[hb!]
\begin{center}
\vspace{8cm}
\includegraphics{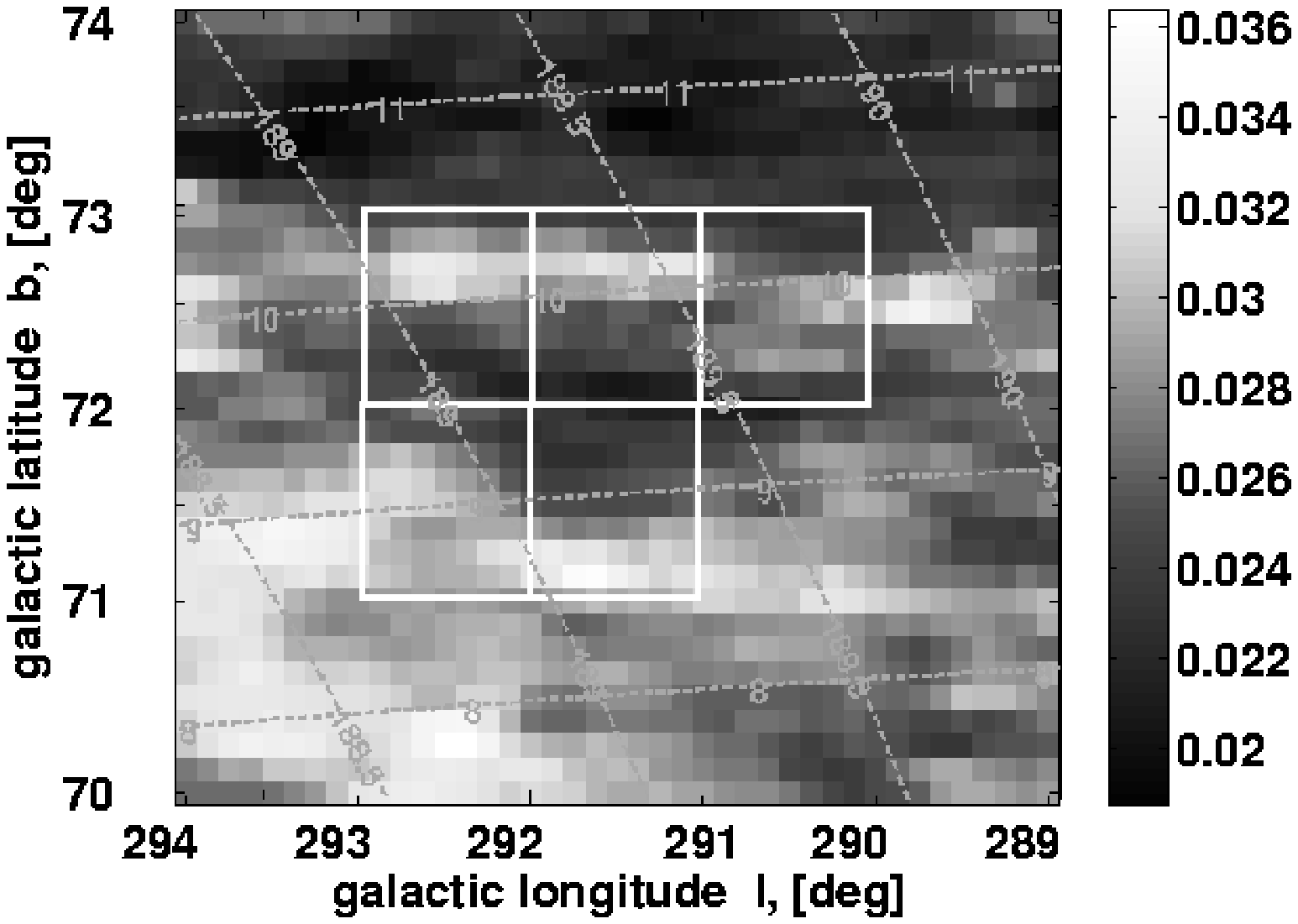}
\end{center}
\caption{%
SFD $E_{B - V}$ reddening map for the first region. White rectangles denote the one-square degree cells of the first region, grey dashed lines denote the equatorial coordinate grid. The values of the $E_{B - V}$ for each ``pixel'' are according to the grey scale at the right.
\label{Fig:1area_SFD}}
\end{figure*}

\newpage
\begin{table*}[ht!]
\caption{Extinction calculated for the first extinguished region vs. that predicted by the SFD map \label{Table:1region}}
\begin{center}
\begin{tabular}{c|c|c|c}

\hline
{Band} & {Direct } & { Model} &  { SFD} \\
\hline
    $A_{u'}$ & $0.^m16\pm0.^m05$ & $0.^m19^{+0.05}_{-0.07}$ & $0.^m14\pm0.^m02$\\
    $A_{g'}$ & $0.^m24\pm0.^m05$ & $0.^m34^{+0.06}_{-0.06}$ & $0.^m10\pm0.^m02$\\
    $A_{r'}$ & $0.^m26\pm0.^m09$ & $0.^m28^{+0.05}_{-0.03}$ & $0.^m07\pm0.^m01$\\
    $A_{i'}$ & $0.^m28\pm0.^m07$ & $0.^m37^{+0.06}_{-0.04}$ & $0.^m05\pm0.^m01$\\
    $A_{z'}$ & $0.^m29\pm0.^m06$ & $0.^m50^{+0.05}_{-0.07}$ & $0.^m04\pm0.^m01$\\
\hline
\end{tabular}  
\end{center}
\end{table*}

\begin{figure*}[ht!]
\begin{center}
\vspace{7.5cm}
\includegraphics{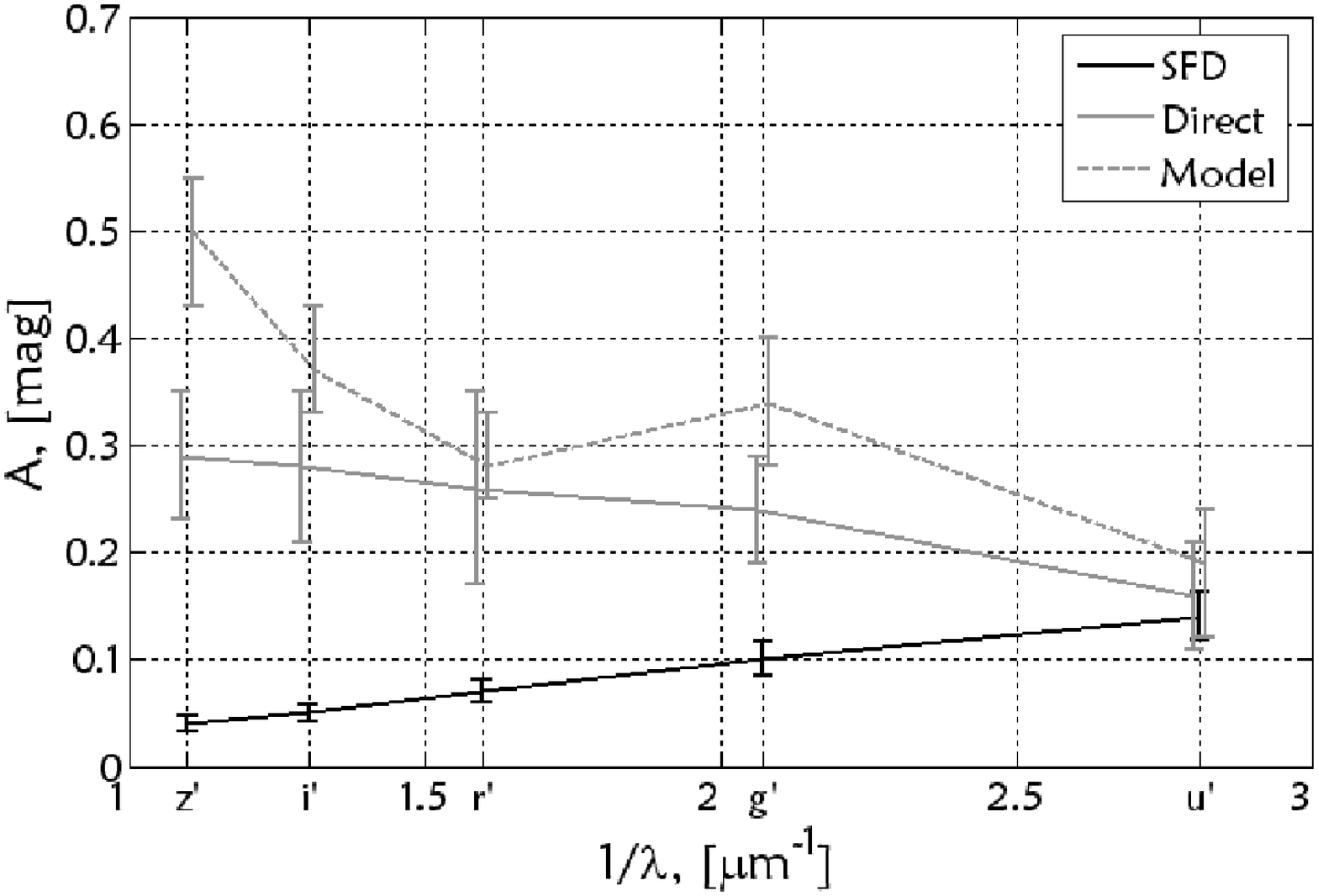}
\end{center}
\caption{%
``Extinction law'' for the first extinguished region predicted by the SFD map, calculated directly and calculated from model. 
\label{Fig:1area_extlaw}}
\end{figure*}

The region is centered on $(l, b) = (292^\circ, 72^\circ)$ or $(\alpha, \delta) = (12.^h6, 9.^\circ5)$ and covers $\sim$1.5 square degrees. Figure \ref{Fig:1area_SFD} shows the detailed SDF map of the region and demonstrates that the extinguished patch does not coincide with an enhanced FIR emission region. The Wolf diagrams indicate a $g'$ band extinction of about $0.^m24$. The specific values, derived directly from the Wolf diagrams or via the thin screen model, are given in Table \ref{Table:1region}. We found that the distance to the extinguishing medium is $890^{+90}_{-80}$ pc using the direct method, or $1510^{+60}_{-60}$ pc using the thin screen model. We note that these distance estimates are very different and discuss this in Section 4. Using the different extinction values from Table \ref{Table:1region} we plotted the wavelength dependence of the extinction in the SDSS bands, which is shown in Figure \ref{Fig:1area_extlaw}.  

\subsection{Second extinguished region}

\begin{figure*}[ht!]
\begin{center}
\vspace{9cm}
\includegraphics{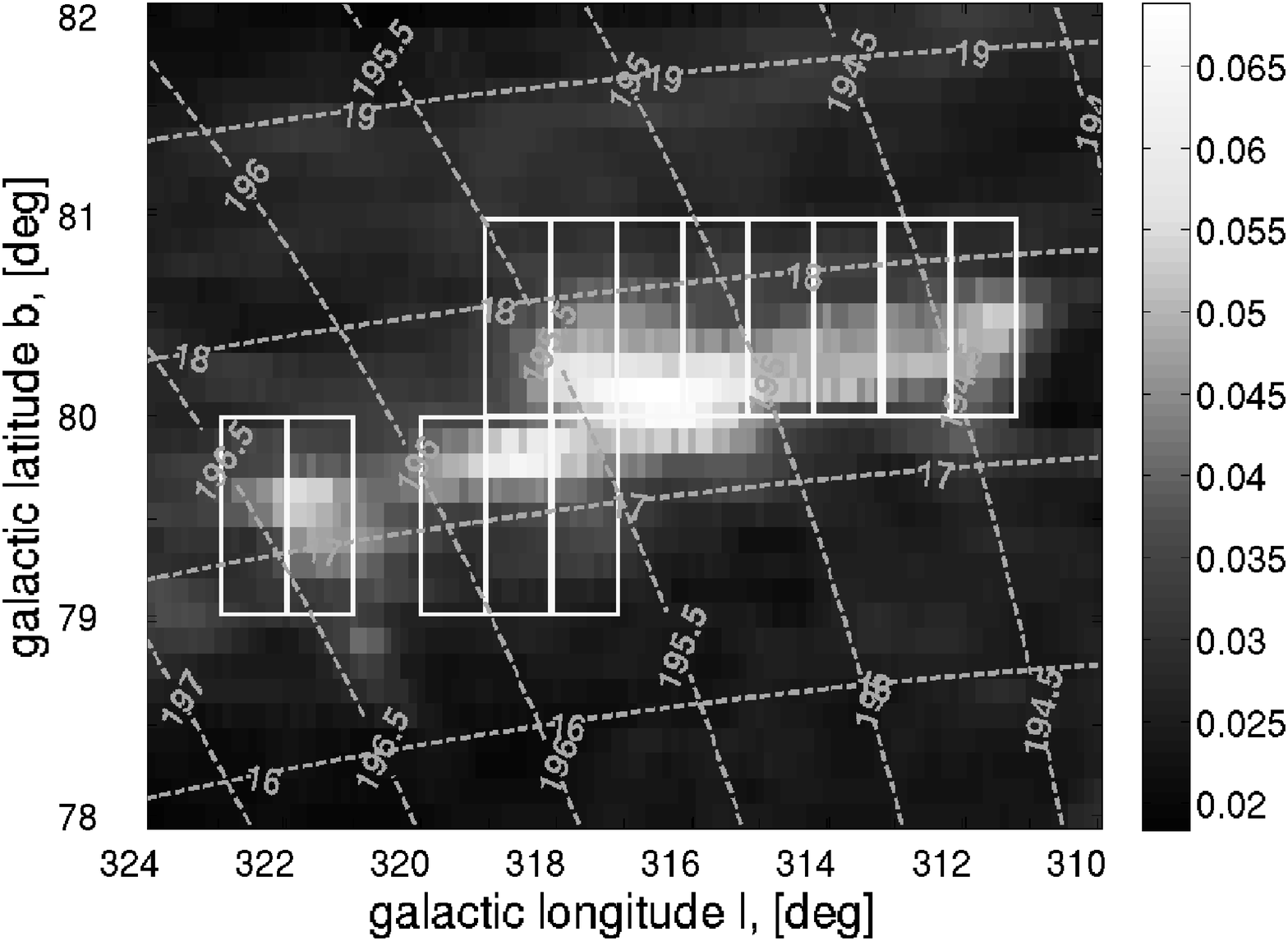}
\end{center}
\caption{%
SFD $E_{B - V}$ reddening map for the second region. White rectangles denote the one-square degree cells of the second region, grey dashed lines denote the equatorial coordinate grid. $E_{B - V}$ scale at the right.
\label{Fig:2area_SFD}}
\end{figure*}

\newpage
\begin{table*}[ht!]
\caption{Extinction calculated for the second extinguished region vs. that predicted by the SFD map \label{Table:2region}}
\begin{center}
\begin{tabular}{c|c|c|c}
\hline
{Band} & {Direct } & { Model} &  { SFD} \\
\hline
    $A_{u'}$ & $0.^m22\pm0.^m02$ & $0.^m47^{+0.28}_{-0.15}$ & $0.^m21\pm0.^m03$\\
    $A_{g'}$ & $0.^m26\pm0.^m02$ & $0.^m42^{+0.06}_{-0.08}$ & $0.^m15\pm0.^m03$\\
    $A_{r'}$ & $0.^m26\pm0.^m03$ & $0.^m41^{+0.06}_{-0.09}$ & $0.^m11\pm0.^m02$\\
    $A_{i'}$ & $0.^m25\pm0.^m03$ & $0.^m31^{+0.08}_{-0.05}$ & $0.^m09\pm0.^m01$\\
    $A_{z'}$ & $0.^m29\pm0.^m08$ & $0.^m35^{+0.08}_{-0.09}$ & $0.^m06\pm0.^m01$\\
\hline
\end{tabular}  
\end{center}
\end{table*}

\begin{figure*}[ht!]
\begin{center}
\vspace{7.5cm}
\includegraphics{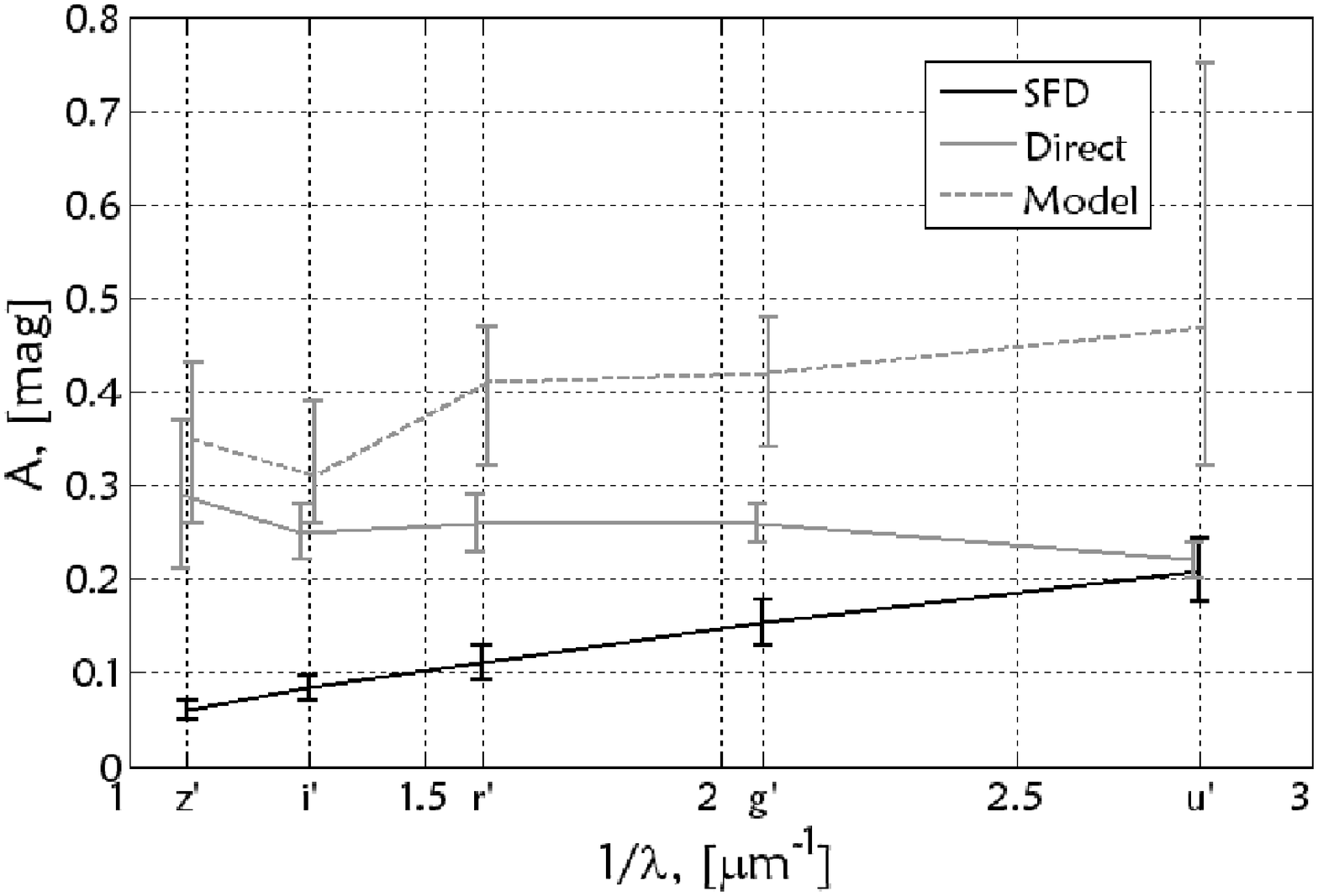}
\end{center}
\caption{%
``Extinction law'' for the second extinguished region predicted by the SFD map, calculated directly and calculated from model. 
\label{Fig:2area_extlaw}}
\end{figure*}

The region is centered on $(l, b) = (316^\circ, 80^\circ)$ or $(\alpha, \delta) = (13.^h0, 17.^\circ5)$ and covers $\sim$2.3 square degrees. Figure \ref{Fig:2area_SFD} shows the detailed SDF map of the region from which it appears that, in this case, the extinguishing medium coincides with a higher FIR emission patch. The Wolf diagrams indicate a $g'$ band extinction of about $0.^m26$. The specific values derived directly from the Wolf diagrams or with the thin screen model are given in Table \ref{Table:2region}. The distance to the extinguishing medium is $1600^{+80}_{-70}$ pc using the direct method, or $1920^{+370}_{-370}$ pc using the thin screen model. Using the different extinction values from Table \ref{Table:2region}, we plotted the wavelength dependence of the extinction in the SDSS bands, which is shown in Figure \ref{Fig:2area_extlaw}.  

\subsection{Third extinguished region}

\begin{figure*}[ht!]
\begin{center}
\vspace{9cm}
\includegraphics{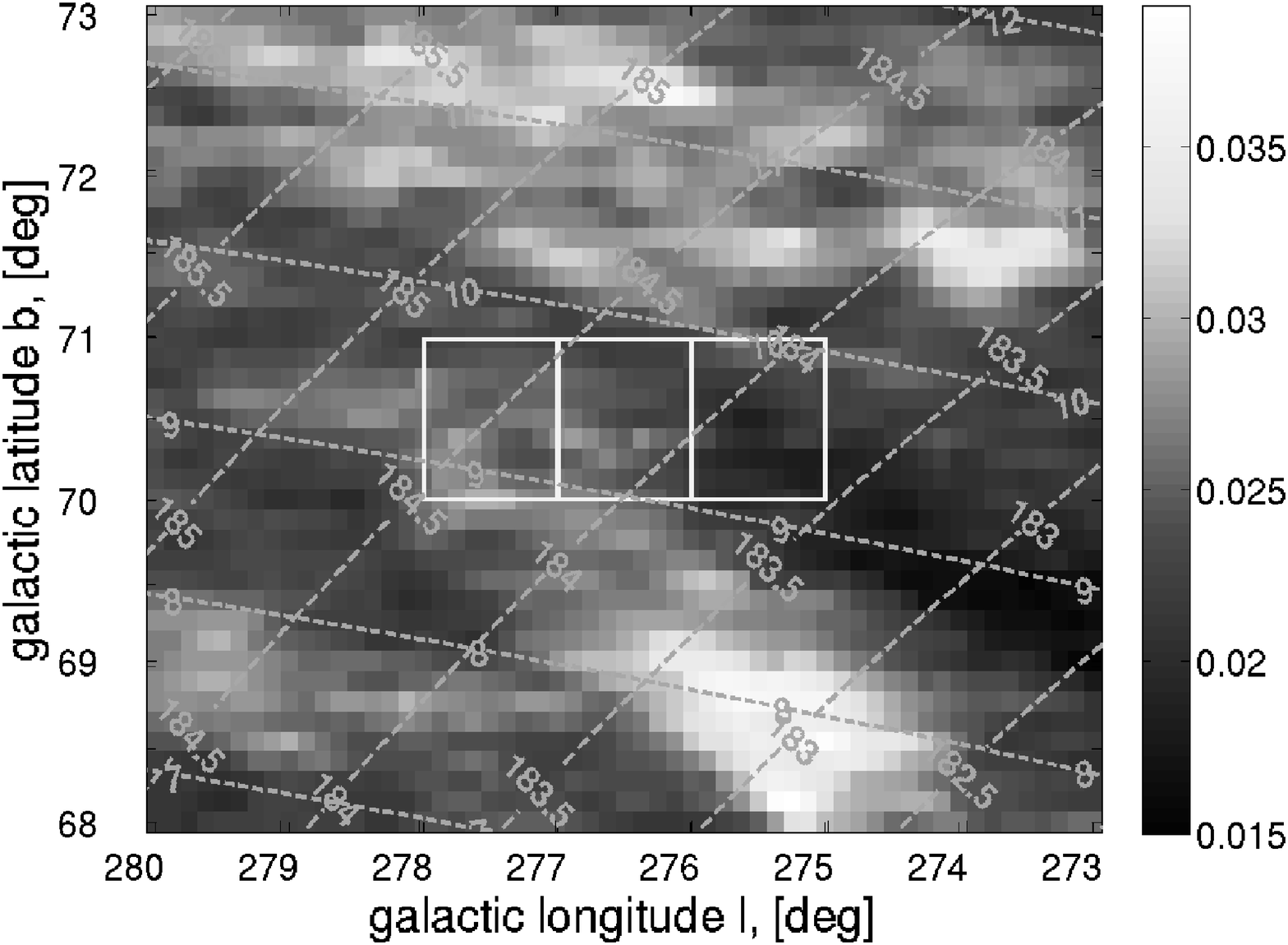}
\end{center}
\caption{%
SFD $E_{B - V}$ reddening map for the third region. White rectangles denote the one-square degree cells of the first region, grey dashed lines denote the equatorial coordinate grid. $E_{B - V}$ scale at the right.
\label{Fig:3area_SFD}}
\end{figure*}

\newpage
\begin{table*}[ht!]
\caption{Extinction calculated for the third extinguished region vs. that predicted by the SFD map \label{Table:3region}}
\begin{center}
\begin{tabular}{c|c|c|c}
\hline
{Band} & {Direct } & { Model} &  { SFD} \\
\hline
    $A_{u'}$ & $0.^m26\pm0.^m04$ & $0.^m25^{+0.10}_{-0.06}$ & $0.^m12\pm0.^m02$\\
    $A_{g'}$ & $0.^m29\pm0.^m05$ & $0.^m40^{+0.10}_{-0.06}$ & $0.^m09\pm0.^m01$\\
    $A_{r'}$ & $0.^m26\pm0.^m07$ & $0.^m36^{+0.05}_{-0.05}$ & $0.^m06\pm0.^m01$\\
    $A_{i'}$ & $0.^m27\pm0.^m07$ & $0.^m41^{+0.08}_{-0.08}$ & $0.^m05\pm0.^m01$\\
    $A_{z'}$ & $0.^m28\pm0.^m05$ & $0.^m50^{+0.15}_{-0.12}$ & $0.^m03\pm0.^m01$\\
\hline
\end{tabular}  
\end{center}
\end{table*}

\begin{figure*}[ht!]
\begin{center}
\vspace{7.5cm}
\includegraphics{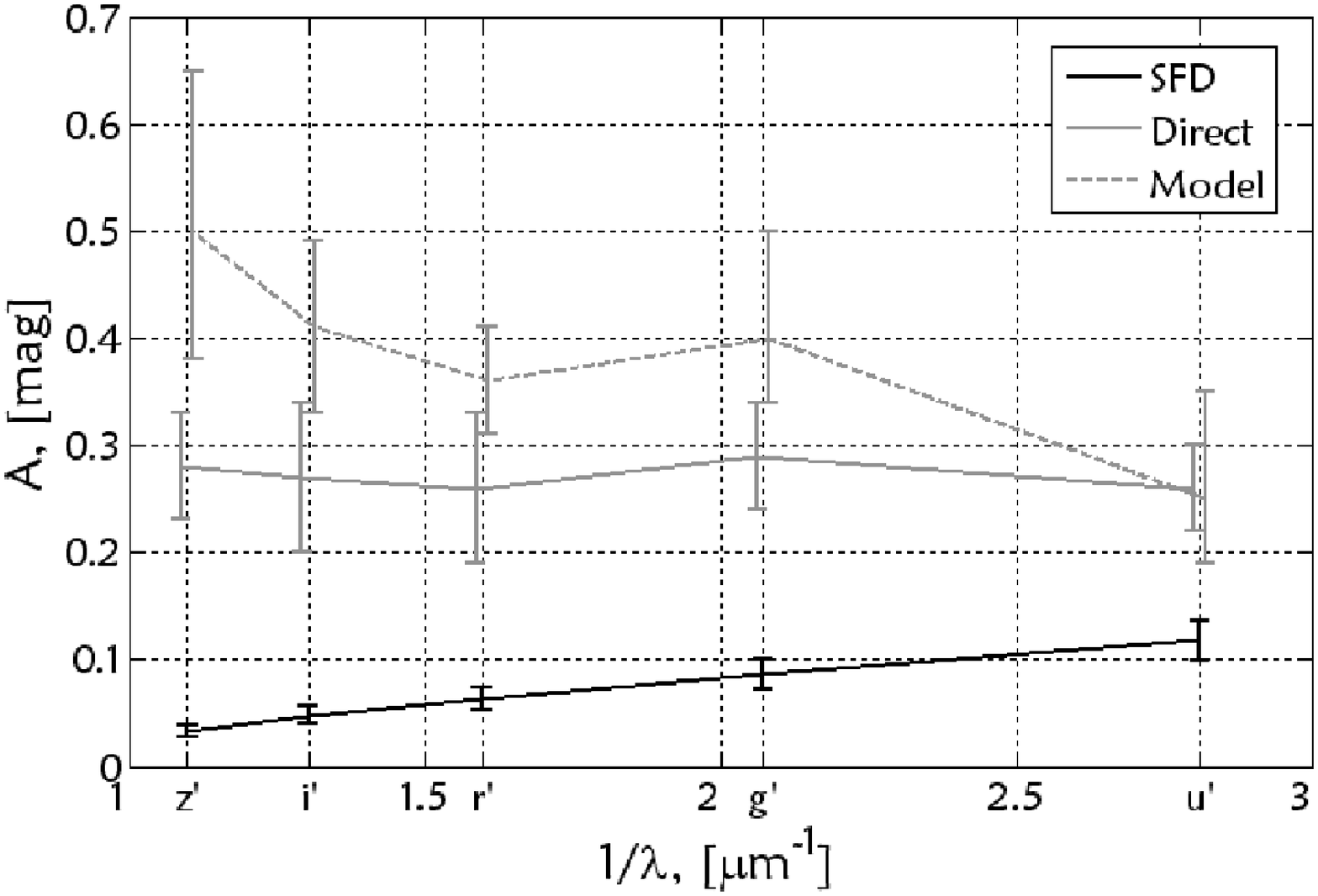}
\end{center}
\caption{%
``Extinction laws'' for the third extinguished region predicted by the SFD map, calculated directly and calculated from model. 
\label{Fig:3area_extlaw}}
\end{figure*}

The region is centered on $(l, b) = (276.^\circ5, 70.^\circ5)$ or $(\alpha, \delta) = (12.^h3, 9.^\circ5)$ and covers $\sim$1.0 square degrees. Figure \ref{Fig:3area_SFD} shows the detailed SDF map of the  region. Here, as in the first extinguished region, the FIR emission is at a local minimum. The Wolf diagrams indicate a $g'$ band extinction of about $0.^m29$. The specific values derived directly from the Wolf diagrams or using the thin screen model are given in Table \ref{Table:3region}. The distance to the extinguishing medium is $800^{+60}_{-50}$ pc using the direct method, or $1320^{+50}_{-30}$ pc using the thin screen model. Here also the distance estimates are very different, as found for the first extinguished region. This is relegated to the discussion in Section 4. Using the different extinction values from Table \ref{Table:3region}, we plotted the wavelength dependence of the extinction in the SDSS bands, which is shown in Figure \ref{Fig:3area_extlaw}.  

\section{Discussion and conclusions}

We presented here a method to derive MW extinction using SDSS stellar photometry analyzed with Wolf diagrams. We showed that this method can not only estimate the amount of extinction in all five bands, but can also estimate a distance to the extinguishing medium, if this is assumed to be a compact and well-defined dust cloud. 

We applied the method to high galactic latitude areas predicted by SFD to have significant extinction, but we found none. We confirmed that in the direction of three HST KP galaxies there is no significant extinction.

We scanned the entire $\sim$650 square degree area and found three regions with measurable extinction. We determined extinction and distance to the extinguishing medium by two alternative methods: direct measurement using the Wolf diagrams, and fitting a thin uniform dust screen model.
The direct measurements produced significantly smaller distances to the extinguishing sources in all the regions when compared with those from the model fitting, while the extinction values estimated by the two methods are comparable. We suggest that the distances estimated by the model fitting should be adopted, since in the direct measurements we excluded $\sim40\%$ of the stars because they were very red or had non-main-sequence colour relations or were fainter than the SDSS limiting magnitudes, and also because of the somewhat uncertain estimation of the SP.

We also checked the effect of possible biases on our estimates. The distances to the extinguishing medium in the three regions may be underestimated by less than 3\% due to the Malmquist bias. The effects of tiling, substructure and extremely red stars rejection were found to be negligible. The binarity bias could, in principal, produce an error in the distance estimation, but the extinction measurement, as performed here, would not be affected. We did not treat the binarity bias in our study, since it is very complicated and is controversial.

We checked the effect of different median longitudes and latitudes of the analyzed regions and the reference fields on the extinction estimation.
The differences between the median longitudes and latitudes of the analyzed regions and the reference fields
were of order $0^{\circ}.2$. The effect of different centers was found to be insignificant, since its contribution
to the extinction estimation is about $\sim$0.01 mag for all the bands in all the regions.

We also performed the following 'sanity' check: we split the reference field into two smaller and concentric reference 
fields and compared the distributions of the analyzed region with that of the first reference field and those of the two reference fields. In regions A, B and C the reduced $\chi^2$ value was less than 2 for the comparison between the 
distributions of the analyzed region and of the first reference field and for the comparison between the distributions of 
the two reference fields. In regions 1, 2 and 3 the reduced $\chi^2$ value was less than 2 for the comparison between the 
two reference fields, while the reduced $\chi^2$ value for the comparison between the distributions of 
the analyzed region and of the first reference field was larger than 10. This shows that the analyzed region differs from its surroundings, while the first reference field is very similar to the second reference field.

We performed the other sanity check by breaking the extinguished region into two smaller areas and comparing their 
star count distributions to the distribution of the unchanged reference field. The distributions of the smaller regions were similar
and the extinctions derived from them were consistent with those of the entire analyzed regions.

The three extinguishing clouds found here are a few tens of pc wide, and thus are not associated with circumstellar disks. The clouds produce 0.2--0.4 mag of $g'$-band extinction, and are located 1--2 kpc away or 0.5--1 kpc above the galactic plane.  
The extinction measured in all three regions does not follow the accepted MW wavelength dependence. The trend of the extinction law from the direct calculation and from the model fitting remains rather constant with wavelength, indicating grey extinction from large dust grains. This is the first clear evidence of grey extinction in the MW interstellar medium at high galactic latitudes.

The results for the second extinguished region differ qualitatively from the results for the first and the third region and the distances estimated by two methods for it are similar. In the second extinguished region the SFD map shows stronger emission than in the surroundings, while in the first and the third region the emission is weaker than in the surroundings. The extinction law estimated by the model fitting and presented in Figure \ref{Fig:2area_extlaw} shows a rising trend, although this is weaker than the mean MW extinction law trend. The extinction in this region is probably not grey, but weakly wavelength-dependent (greyish). 

The distances and extinction curves for the first and the third regions are very similar. The angular distance between these regions is $15^{\circ}$ and their galactic latitude is $b\approx70^{\circ}$ (see Figure \ref{Fig:ext_all}). Adopting a mean distance of 1400 pc, we estimate the projected distance between the clouds to be $\sim130$ pc, one order of magnitude smaller than the distance from the observer to these regions. We conclude that the dust clouds in these regions may possibly have the same origin and belong to the same dust cloud complex.

All three extinguished regions found in this study show wavelenght-independent or weakly wavelength-dependent extinction, are located more than 1 kpc above the galactic plane, and all are in the third quadrant. It is tempting to connect them with other strange entities in this part of the sky, such as possible cannibalized dwarf galaxies, but with the information collected so far this would be speculative.

However, our finding of significant extinction at high galactic latitude, and the determination that this is grey extinction, should at the least inject a note of caution in studies of high-latitude objects and extragalactic sources. Milky Way dust can affect observational results in subtle and unexpected manners.

\section*{Acknowledgments}
Funding for the SDSS and SDSS-II has been provided by the Alfred P. Sloan Foundation, the Participating Institutions, the National Science Foundation, the U.S. Department of Energy, the National Aeronautics and Space Administration, the Japanese Monbukagakusho, the Max Planck Society, and the Higher Education Funding Council for England. The SDSS Web Site is http://www.sdss.org/.

The SDSS is managed by the Astrophysical Research Consortium for the Participating Institutions. The Participating Institutions are the American Museum of Natural History, Astrophysical Institute Potsdam, University of Basel, University of Cambridge, Case Western Reserve University, University of Chicago, Drexel University, Fermilab, the Institute for Advanced Study, the Japan Participation Group, Johns Hopkins University, the Joint Institute for Nuclear Astrophysics, the Kavli Institute for Particle Astrophysics and Cosmology, the Korean Scientist Group, the Chinese Academy of Sciences (LAMOST), Los Alamos National Laboratory, the Max-Planck-Institute for Astronomy (MPIA), the Max-Planck-Institute for Astrophysics (MPA), New Mexico State University, Ohio State University, University of Pittsburgh, University of Portsmouth, Princeton University, the United States Naval Observatory, and the University of Washington.

We would like to express our gratitude to Ido Finkelman for valuable advices and pleasant conversation and to an anonymous referee whose remarks improved the paper.

\end{document}